\documentclass[10pt,a4paper]{article}
\usepackage[applemac]{inputenc}
\usepackage[T1]{fontenc}
\usepackage{mathtools}
\usepackage{authblk}
\usepackage{amsmath}
\usepackage{amssymb}
\usepackage{stmaryrd}
\usepackage{hyperref}
\usepackage{xcolor}
\usepackage[symbol]{footmisc}

\newcommand{\Rea}{\mathbb{R}}

\newcommand{\Id}{\hat{\mathbb{I}}}

\newcommand{\Hi}{\mathcal{H}}

\newcommand{\Nullo}{\mathbb{O}}

\newcommand{\E}{\mathcal{E}}

\newcommand{\Ex}{\mathbb{E}}
\newcommand{\Ito}{It\^{o} }
\newcommand{\Tr}[1]{\mbox{Tr}\left[#1\right]}

\newcommand{\PS}{(\Omega,\E,P)}

\newcommand{\Slog}{\mathcal{L}og}

\title{Unitary time-evolution in stochastic time-dependent Hilbert spaces}
\author[1,2]{Curcuraci L.}
\author[1]{Bacchi S.}
\author[1,2]{Bassi A.}
\affil[1]{Department of Physics, University of Trieste, Strada Costiera 11 34151, Trieste, Italy}
\affil[2]{Istituto Nazionale di Fisica Nucleare, Trieste Section, Via Valerio 2 34127, Trieste, Italy}
\begin{document}
	
	\maketitle
	\begin{abstract}
		In this work we study the unitary time-evolutions of quantum systems defined on infinite-dimensional separable time-dependent Hilbert spaces. Two possible cases are considered: a quantum system defined on a stochastic interval and another one defined on a Hilbert space with stochastic integration measure (stochastic time-dependent scalar product). The formulations of the two problems and a comparison with the general theory of open quantum systems are discussed. Possible physical applications of the situations considered are analyzed.
	\end{abstract}

	\section{Introduction}
	
	Suppose we have a quantum system whose states are represented by vectors of a time-dependent Hilbert space $\Hi_t$. At each time $t$, its state is a vector $|\psi_t\rangle \in \Hi_t$ and its observables are operators defined on $\Hi_t$. In this article we will deal with the problem of unitary evolution of the quantum state in situations in which the time-evolution of $\Hi_t$ is regulated by stochastic differential equations. To fix the ideas, we will consider  separable infinite-dimensional Hilbert spaces like $L_2(\mathcal{M},dm)$, where $\mathcal{M}$ is some set and $dm$ some measure on it. Two are the possible time-dependences considered: the time-dependence of the set, i.e. $L_2(\mathcal{M}_t, dm)$, and time-dependence of the measure, i.e. $L_2(\mathcal{M},dm_t)$. Many physical situations fall in these two categories. A particle trapped in a vibrating well is an example of situations in which an Hilbert space $L_2(\mathcal{M}_t,dm)$ can be used. Another interesting physical situation can be the case of a particle moving in a vibrating space, which is an example of Hilbert space $L_2(\mathcal{M},dm_t)$. This last case is particularly interesting since the recent observations of gravitational waves \cite{abbott2016observation,abbott2017gw170817}, opens the possibility of detecting the "cosmic gravitational wave background", which may be effectively considered as a stochastic process of the metric testable in high-precision quantum experiments.
	
	The case $L_2(\mathcal{M}_t,dm)$ is well studied for the case of deterministic time-evolutions. The case when $\mathcal{M}_t$ is a 1D interval with one moving extremum is studied in \cite{munier1981schrodinger}, while a more systematic approach is presented in \cite{di2013quantum}. The 2D and 3D cases are discussed in \cite{di2015quantum} and in \cite{facchi2008quantum} the connection of this problem with the quantum Zeno dynamics is presented. In section \ref{2} we analyze the more general case of a stochastic interval, discussing its physical interpretation and the connection (if any) with the general theory of open quantum system. Regarding the $L_2(\mathcal{M},dm_t)$ case, the deterministic case is well studied, especially for its connection with gravity \cite{dewitt1952point, dewitt1957dynamical, da1981quantum}. The dynamics in the case of a generic time-dependent Hilbert space is discussed in \cite{mostafazadeh2004time}, however the time-evolution of the measure is derived by requiring unitarity, rather than imposing it a priori (which is the case considered here). In addition, only the finite dimensional case is considered. In section \ref{3} we construct the Hilbert space of a quantum system on a Hilbert space with stochastic integration measure. Doing that, the unitary dynamics is introduced and different concrete representations of it are discussed.
	
	\section{Unitary dynamics on an interval with diffusive boundary }\label{2}
	
	In this first part we consider the case of a quantum system with Hilbert space $L_2([a_t,b_t],dx)$, where $a_t$ and $b_t$ change randomly in time. The method we will use is based on defining a map from $L_2([a_t,b_t],dx)$ to the time-independent Hilbert space $L_2(\left[ -\frac{1}{2}, \frac{1}{2}\right],dx)$ in order to define a time-derivative and consequently to compute the time-evolution equation. Since the measure $dx$ of the Hilbert space does not change, we will use simply the symbol $L_2([a_t,b_t])$ if no confusion arises. This procedure generalizes straightforwardly to the 3D case, i.e. situations like $\mathcal{M}_t = [a^x_t,b^x_t] \times [a^y_t,b^y_t] \times [a^z_t,b^z_t]$, despite calculations may become more involved.
	
	\subsection{Mapping on a time independent Hilbert space}\label{2.1}
	
	Let us consider a quantum particle confined in a randomly changing interval of $\Rea$. More precisely, let $[a_t,b_t]$ be such an interval and assume that $a_t$ and $b_t$ are two independent diffusion processes. More specifically, let $(\Omega,\E,P)$ be a probability space and chose a filtration $\{\mathcal{F}_t\}_{t \in [0,T]}$ on it. The two extrema of the interval, $a_t(\omega)$ and $b_t(\omega)$, are two diffusion processes adapted to the chosen filtration. Hence, the SDEs describing their time evolution are
	\begin{equation}\label{BoundarySDE}
	\begin{cases}
	d a_t(\omega) = \mu_a(a_t(\omega),t)dt + \sigma_a(a_t(\omega),t)dW_t^{a}(\omega) \\
	d b_t(\omega) = \mu_b(b_t(\omega),t)dt + \sigma_b(b_t(\omega),t)dW_t^{b}(\omega)
	\end{cases}
	\end{equation}
	where $\mu_i$ and $\sigma_i$ are respectively the drift and variance functions of the processes, for any $i = a,b$. The two Wiener processes $W_t^{a}(\omega)$ and $W_t^b(\omega)$ are assumed to be standard and independent. In what follows, the dependence on $\omega$ of the processes is explicitly displayed only when necessary. The Hilbert space describing the quantum particle is $L_2([a_t,b_t])$, which is stochastic because of the interval on which it is defined. We assume that the particle's Hamiltonian is 
	\begin{equation}\label{Hamiltonian}
	\hat{H} = \hat{T} + \hat{V}
	\end{equation}
	 where
	\begin{equation}\label{Kinetic_term}
	\hat{T} = -\frac{\hat{P}^2}{2m},
	\end{equation}
	($m \in \Rea^+$ is a constant and $\hat{P}$ the momentum operator) is the kinetic term and $\hat{V}$ is a suitable $\| \hat{T} \|$-bounded operator (in the sense of the Kato-Rellich theorem \cite{moretti2013spectral}) representing a potential. We assume that $\hat{H}$ is at least self-adjoint on its domain $D(\hat{H}) \subset L_2([a_t,b_t])$.  By the Kato-Rellich theorem, the self-adjointness  of $\hat{T}$ it is sufficient to ensure the self-adjointness of $\hat{H}$. Later we will discuss various possible domains for $\hat{H}$. In particular we will consider domains of the form
	\begin{equation*}
		\mathcal{D}(\hat{H}) = \{ \psi \in H^2\left([a_t,b_t]\right) | \mbox{boundary conditions in $a_t$ and $b_t$} \},
	\end{equation*}
	and so the choice of a domain over another reduces to choice of the boundary conditions. Note that these boundary conditions are time-dependent. The particular time dependency of this Hilbert space makes any time-evolution on it ill-defined: it is not possible to define, in a straightforward manner, the time derivative of the state vector using the standard definition, as discussed in \cite{di2013quantum}. Moreover, even considering this problem as purely formal, the ordinary methods for the solution of differential equations do not work in presence of time dependent boundary conditions \cite{munier1981schrodinger}. To overcome these difficulties, one observes that $[a_t,b_t]\subset \Rea$, which means that our particle can be described on (a subspace of) the Hilbert space $L_2(\Rea)$. For later convenience, one rewrites the interval $[a_t, b_t]$ as
	\begin{equation}\label{Newinterval}
	I_{m_t,l_t} := \left[m_t - \frac{l_t}{2}L_0, m_t + \frac{l_t}{2}L_0\right]
	\end{equation}
	where $m_t := (a_t + b_t)/2$ and $l_t = |a_t - b_t|/L_0$. $L_0$ is a fixed positive quantity needed in order to make $l_t$ dimensionless\footnote{Note that $l_t\geqslant0$ by construction. This is done because using stochastic processes for the evolution of the boundary, one cannot in general guarantee that the difference of the two extrema does not change sign. This is a very important fact as will be clear later.}. Hereafter, we set $L_0 = 1$ without lose any generality. Starting from \eqref{BoundarySDE}, the SDE fulfilled by $m_t$ and $l_t$ can be computed. From a trivial application of the \Ito formula, one can easily derive that $m_t$ fulfills the SDE
	\begin{equation}\label{meanSDE}
	d m_t = \mu_1(t)dt + dX^1_t,
	\end{equation}
	where 
	\begin{equation*}
	\mu_1(t):= \frac{\mu_a(t) + \mu_b(t)}{2}
	\end{equation*}
	and
	\begin{equation*}
	dX_t^1 := \frac{\sigma_a(t)dW_t^a + \sigma_b(t)dW_t^b}{2}.
	\end{equation*}
	More care is needed in the derivation of the SDE for $l_t$. In fact, $f(x) = |x|$ is not a $C^2(\Rea)$ function, as required by the \Ito formula. However, one can apply the Tanaka formula \cite{klebaner2005introduction,protter2005stochastic,bjork2015pedestrian} from which one obtains
	\begin{equation*}
	d l_t = \mbox{sign}(a_t - b_t)d(a_t - b_t) + \delta(a_t - b_t) d \llbracket a_t-b_t , a_t - b_t \rrbracket,
	\end{equation*}
	where $\delta(x)$ is the usual Dirac delta function. Above, the brackets $ \llbracket , \rrbracket$ are the quadratic covariation brackets for a stochastic process typically used in stochastic calculus \cite{klebaner2005introduction}.  From \eqref{BoundarySDE} one has $ d\llbracket a_t-b_t , a_t - b_t \rrbracket = (\sigma_a^2 + \sigma_b^2)dt$, thus
	\begin{equation}
	d l_t = \tilde{\mu}_2(t)dt + dX_t^2
	\end{equation}
	where 
	\begin{equation*}
	\tilde{\mu}_2(t):= \mbox{sign}(a_t - b_t)(\mu_a(t) - \mu_b(t)) + (\sigma_a^2(t) + \sigma_b^2(t)) \delta(a_t - b_t),
	\end{equation*}
	and
	\begin{equation*}
	dX_t^2 := \mbox{sign}(a_t - b_t) (\sigma_a(t)dW_t^a - \sigma_b(t)dW_t^b).
	\end{equation*}
	Note that  the Dirac delta function in the drift term contributes only when $l_t = 0$. As it will be clear later, $l_t=0$ is problematic both from the physical and mathematical point of view. However, we will see in the examples how to avoid this kind of situation by properly describing the motion of the boundaries. For later convenience, we also derive the stochastic differential for $\log l_t$, which is
	\begin{equation}\label{logSDE}
	d \log l_t = \mu_2(t) dt +  \frac{1}{l_t}dX_t^2
	\end{equation}
	with
	\begin{equation*}
	\mu_2(t) :=  \frac{\tilde{\mu}_2(t)}{l_t}  - \frac{\sigma_a^2(t) + \sigma_b^2(t)}{2l_t^2}.
	\end{equation*}
    Since $l_t \geqslant 0$ by construction, the above SDE is always well defined except when $l_t=0$, where $\log l_t$ diverges and its SDE is ill defined. We will come back on this point later.
	
	In general, we can write $L_2(I_{m_t,l_t}) \subset L_2(I_{m_t,l_t}) \oplus L_2(I_{m_t,l_t}^c) = L_2(\Rea)$ where by $I_{m_t,l_t}^c$ we mean the complementary set of  $I_{m_t,l_t}$ with respect to $\Rea$. Because of that, the particle can be described by an extended wave function $\Psi = \psi + \phi$: $\psi \in L_2(I_{m_t,l_t})$ is the particle's wave-function, while $\phi \in L_2(I_{m_t,l_t}^c)$ is an arbitrary function. The time-evolution is obtained by using the extended Hamiltonian
	\begin{equation}\label{ExtendedHamiltonian}
	\hat{H}_{ext}(m_t,l_t) := \hat{H} \oplus_{m_t,l_t} \hat{\Nullo}
	\end{equation}
	and using the ordinary time derivative defined in $L_2(\Rea)$. Above, the symbol $\oplus_{m_t,l_t}$ is used to emphasize that this direct sum of operators is time dependent, because the two operators are defined on two time dependent Hilbert spaces \cite{di2013quantum}. Because of this $\hat{H}_{ext}(m_t,l_t)$ is a stochastic operator since it depends on $m_t$ and $l_t$. In particular, the domain of $\hat{H}_{ext}(m_t,l_t)$, $\mathcal{D}(\hat{H}_{ext}(m_t,l_t)) = \mathcal{D}(\hat{H})\oplus L_2(I_{m_t,l_t}^c)$, is time-dependent and stochastic. The common way to tackle this kind of problems, is to describe the time evolution on a Hilbert space where the domain is fixed \cite{di2013quantum}. The interval \eqref{Newinterval} can be mapped into a fixed interval in an easy way: taken $x \in I_{m_t,l_t}$ one first performs a translation $x \rightarrow x - m_t$ and then a rescaling $x \rightarrow x/l_t$. In this way, one maps $I_{m_t,l_t}$ into the interval $\left[ -\frac{1}{2}, \frac{1}{2} \right]$, i.e. the Hilbert space $L_2(I_{m_t,l_t}) \oplus L_2(I_{m_t,l_t}^c)$ is mapped into $L_2(\left[ -\frac{1}{2}, \frac{1}{2}\right]) \oplus L_2(\left[ -\frac{1}{2}, \frac{1}{2}\right]^c)$. This mapping can be constructed in the following way. Let $\hat{X}$ and $\hat{P}$ be the ordinary position and momentum operators on $L_2(\Rea)$, the operator
	\begin{equation*}
	\hat{G} = \frac{\hat{X} \hat{P} + \hat{P} \hat{X}}{2}
	\end{equation*}
	is the so called dilation operator \cite{ballentine1998quantum}. From the prescription given, the mapping is done by using the operator $\hat{W}(m_t,l_t) : L_2(I_{m_t,l_t}) \oplus L_2(I_{m_t,l_t}^c) \rightarrow L_2(\left[ -\frac{1}{2}, \frac{1}{2}\right]) \oplus L_2(\left[ -\frac{1}{2}, \frac{1}{2}\right]^c)$ defined as
	\begin{equation}\label{Transformation}
	\hat{W}(m_t,l_t) := \hat{D}(-\log l_t) \hat{T}(-m_t)
	\end{equation}
	where $T(a) = \exp( -i a \hat{P} )$ is the generator of the spatial translation, $\hat{T}(a)|x\rangle = |x + a\rangle$,  and $\hat{D}(\lambda) = \exp(-i \lambda \hat{G})$ is the generator of the dilation transformation, $\hat{D}(\lambda)|x\rangle = e^{\frac{\lambda}{2}}|e^{\lambda} x \rangle$. Since $\hat{T}(a)$ and $\hat{D}(\lambda)$ are two unitary operators on $L_2(\Rea)$, also $\hat{W}(m_t,l_t)$ is unitary on this Hilbert space. From now on, we omit the dependence on $m_t$ and $l_t$ in $\hat{H}_{ext}(m_t,l_t)$ and $\hat{W}(m_t,l_t)$, if no confusion arises. 
	
	Before to go on with the derivation of the time-evolution, let us discuss the point $l_t =0$ (i.e. $a_t = b_t$). The dilation operator $\hat{D}(-\log l_t)$ in \eqref{Transformation} is ill-defined in this point, since $\log l_t$ diverges. Thus the unitary operator $\hat{W}$ is not defined in this case, meaning that $\hat{W}$ is able to map $ L_2(I_{m_t,l_t}) \oplus L_2(I_{m_t,l_t}^c)$ into $L_2(\left[ -\frac{1}{2}, \frac{1}{2}\right]) \oplus L_2(\left[ -\frac{1}{2}, \frac{1}{2}\right]^c)$ only when $l_t > 0$. Hence the equation we derive in the next section is valid either when $a_t < b_t$ or when $a_t > b_t$but not when $a_t = b_t$. Mathematically this corresponds to having an Hilbert space defined on a single point, i.e. $L_2(\{a_t\},dx)$. It is not difficult to see that this kind of \textquotedblleft Hilbert space\textquotedblright $\mspace{2mu}$would be highly problematic (for example, the scalar product would be aways $0$). Physically $l_t=0$ happens when the two boundaries collide. Clearly this never happens. However starting with two generic diffusion processes, as in \eqref{BoundarySDE}, such situation cannot be avoided in general: only by specifying the SDEs governing the two boundaries one can decide if the boundaries may collide or not. In section \ref{Examples}, we will see how the collision of boundaries can be avoided by using a particular class of diffusion processes called Dyson brownian motion \cite{dyson1962brownian}, which are well defined mathematically and have a very simple physical interpretation. For this reason we assume $l_t > 0$ in the rest of the discussion.

	\subsection{Derivation of the time-evolution equation}
	
	On $L_2(I_{m_t,l_t}) \oplus L_2(I_{m_t,l_t}^c)$, the time-evolution is given by the Schr\"odinger equation with the Hamiltonian \eqref{ExtendedHamiltonian}, i.e.
	\begin{equation}\label{SchEq}
	d|\Psi_t \rangle = -\frac{i}{\hbar}\hat{H}_{ext}|\Psi_t\rangle dt.
	\end{equation}
	Since $\hat{H}_{ext}(m_t,l_t)$ is a stochastic process, the above equation must be interpreted in the sense of stochastic integrals. However we note that the stochastic process $|\Psi_t\rangle$ has finite variation. Indeed, given a continuous differentiable function $g$ with continuous integrable derivative, its variation $V_g(t)$ is given by \cite[Ex 1.5]{klebaner2005introduction}
	\begin{equation*}
	V_g(t) = \int_0^t |g'(s)|ds.
	\end{equation*}
	In our case, the function $t \mapsto |\Psi_t\rangle$ is continuous and differentiable. Its time derivative is continuous and integrable in probability, hence
	\begin{equation*}
	V_{|\Psi\rangle}(t) = \int_0^t \| -\frac{i}{\hbar} \hat{H}_{ext}|\Psi_s \rangle \|ds < + \infty,
	\end{equation*}
	for any $t$ finite. The last inequality follows from the chosen domain for $\mathcal{D}(\hat{H})$ and because $\hat{V}$ is assumed $\| \hat{T} \|$-bounded. By construction, for any $|\Psi_t\rangle \in L_2(I_{m_t,l_t}) \oplus L_2(I_{m_t,l_t}^c)$, the vector 
	\begin{equation}\label{VectorMapping}
	|\Phi_t \rangle = \hat{W}|\Psi_t\rangle
	\end{equation}
	belongs to $L_2(\left[ -\frac{1}{2}, \frac{1}{2}\right]) \oplus L_2(\left[ -\frac{1}{2}, \frac{1}{2}\right]^c)$. Taking the stochastic differential and using \eqref{SchEq}, we get
	\begin{equation*}
	\begin{split}
	d|\Phi_t \rangle &= d(\hat{W}|\Psi_t\rangle) \\
	&= d\hat{W} |\Psi_t\rangle + \hat{W} d|\Psi_t\rangle + d \llbracket \hat{W}, |\Psi_t\rangle \rrbracket \\
	&= d\hat{W} |\Psi_t\rangle +  \frac{1}{i\hbar}\hat{W}\hat{H}_{ext} |\Psi_t\rangle dt + d \llbracket \hat{W}, |\Psi_t\rangle \rrbracket
	\end{split}
	\end{equation*}
	where with the brackets $\llbracket A_t , B_t  \rrbracket$ we refer to the covariation of processes $A_t$ and $B_t$ \cite{klebaner2005introduction}. Since $|\Psi_t\rangle$ is of bounded variation, $\llbracket \hat{W}, |\Psi_t\rangle \rrbracket = 0$, using \eqref{VectorMapping} and the unitarity of $\hat{W}$, we get
	\begin{equation}\label{NewSchrodinger}
	d|\Phi_t \rangle = -\frac{i}{\hbar}\hat{K}|\Phi_t\rangle	 +  d\hat{W}\hat{W}^{\dagger}|\Phi_t\rangle								
	\end{equation}
	where $\hat{K} := \hat{W}\hat{H}_{ext}\hat{W}^{\dagger}$. This equation describes the time evolution of $|\Phi_t\rangle$ on $L_2(\left[ -\frac{1}{2}, \frac{1}{2}\right]) \oplus L_2(\left[ -\frac{1}{2}, \frac{1}{2}\right]^c)$, and is the stochastic analogue of the time evolution equation in \cite{di2013quantum}, due to the presence of the stochastic differential in the last term. From \eqref{Transformation} according to the rule of stochastic calculus we have
	\begin{equation}\label{StochasticDifferentialTerm}
	\begin{split}
	d \hat{W}&= d( \hat{D}(-\log l_t)\hat{T}(-m_t) ) \\
	&= d\hat{D}(-\log l_t)\hat{T}(-m_t) + 	 \hat{D}(-\log l_t)d\hat{T}(-m_t) + d\llbracket  \hat{D}(-\log l_t), \hat{T}(-m_t) \rrbracket,
	\end{split}
	\end{equation}
	and this time the covariation brackets do not vanish in general, as it will be clear in a moment. Using \eqref{meanSDE} and \eqref{logSDE}, one can derive the stochastic differential of $\hat{T}(-m_t) = \exp( im_t\hat{P} )$ and $\hat{D}(-\log l_t) = \exp( i\log l_t \hat{G} )$. After some algebra, one has
	\begin{equation}\label{SDE1}
	d\hat{D}(-\log l_t) = \left[ \frac{i\mu_2(t)}{\hbar} \hat{G}  - \frac{\sigma_a^2(t) + \sigma_b^2(t)}{2\hbar^2l_t^2} \hat{G}^2 \right]\hat{D}(-\log l_t)dt + \frac{i \hat{G}}{\hbar l_t}\hat{D}(-\log l_t)dX_t^2
	\end{equation} 
	and
	\begin{equation}\label{SDE2}
	d\hat{T}(-m_t) = \left[ \frac{i\mu_1(t)}{\hbar} \hat{P}  - \frac{\sigma_a^2(t) + \sigma_b^2(t)}{8\hbar^2} \hat{P}^2 \right]\hat{T}(-m_t)dt + \frac{i \hat{P}}{\hbar}\hat{T}(-m_t)dX_t^1,
	\end{equation}
	where we used $d\llbracket X_1, X_1 \rrbracket = (\sigma_a^2(t) + \sigma_b^2(t))dt/4$ and $d\llbracket X_2, X_2 \rrbracket = (\sigma_a^2(t) + \sigma_b^2(t))dt$.
	Note that the covariation $\llbracket  \hat{D}(-\log l_t), \hat{T}(-m_t) \rrbracket$ does not vanish, as anticipated before. In particular,
	\begin{equation}\label{SDE3}
	\begin{split}
	d\llbracket  \hat{D}(-\log l_t), \hat{T}(-m_t) \rrbracket &= - \frac{1}{\hbar^2 l_t} \hat{G}\hat{D}(-\log l_t)\hat{P}\hat{T}(-m_t) d \llbracket X_t^2,X_t^1 \rrbracket \\
	&= - \mbox{sign}(a_t - b_t)\frac{\sigma_a^2(t) - \sigma_b^2(t)}{2\hbar^2 l_t^2} \hat{G}\hat{P}\hat{D}(-\log l_t)\hat{T}(-m_t)dt
	\end{split}
	\end{equation}
	where we used the unitarity of $\hat{D}(-\log l_t)$, the relation $\hat{D}(-\log l_t) \hat{P} \hat{D}(- \log l_t)^{\dagger} = \hat{P}/l_t$ and $d\llbracket X_1,X_2\rrbracket = \mbox{sign}(a_t - b_t)(\sigma_a^2(t) - \sigma_b^2(t))/2l_t$. From \eqref{SDE1}, \eqref{SDE2} and \eqref{SDE3}, one can rewrite \eqref{StochasticDifferentialTerm} as follows:
	\begin{equation}
	\begin{split}
	d\hat{W} &= \left[ \frac{i\mu_2(t)}{\hbar} \hat{G}  - \frac{\sigma_a^2(t) + \sigma_b^2(t)}{2\hbar^2l_t^2} \hat{G}^2 \right]\hat{W}dt + \frac{i \hat{G}}{\hbar l_t}\hat{W}dX_t^2 \\
	&+ \left[ \frac{i\mu_1(t)}{\hbar l_t} \hat{P}  - \frac{\sigma_a^2(t) + \sigma_b^2(t)}{8\hbar^2l_t^2} \hat{P}^2 \right]\hat{W}dt + \frac{i \hat{P}}{\hbar l_t}\hat{W}dX_t^1 \\
	& - \mbox{sign}(a_t - b_t)\frac{\sigma_a^2(t) - \sigma_b^2(t)}{2\hbar^2 l_t^2} \hat{G}\hat{P} \hat{W} dt
	\end{split}
	\end{equation}
	Inserting this equation in \eqref{NewSchrodinger}, we get
	\begin{equation*}
	\begin{split}
	d|\Phi_t \rangle &= - \frac{i}{\hbar}\hat{K}|\Phi_t\rangle	 +\frac{i}{\hbar} \left[ \frac{\mu_1(t)}{l_t} \hat{P} + \mu_2(t)\hat{G} \right]|\Phi_t\rangle dt \\
	&-  \frac{\sigma_a^2(t) + \sigma_b^2(t)}{2\hbar^2l_t^2}  \left[ \hat{G}^2 + \frac{\hat{P}^2}{8} \right]|\Phi_t\rangle dt 
	- \mbox{sign}(a_t - b_t)\frac{\sigma_a^2(t) - \sigma_b^2(t)}{2\hbar^2 l_t^2} \hat{G}\hat{P}|\Phi_t\rangle dt \\
	&+ \frac{i}{\hbar l_t}\left[ \hat{P}|\Phi_t\rangle dX_t^1 + \hat{G}|\Phi_t\rangle dX_t^2 \right]
	\end{split}
	\end{equation*}
	Using the explicit definitions of $dX_t^1$ and $dX_t^2$, and observing that
	\begin{equation*}
	\mbox{sign}(a_t - b_t)\frac{\hat{P}\hat{G} + \hat{G}\hat{P}}{2} = \mbox{sign}(a_t - b_t)\left[ \hat{G}\hat{P} - \frac{i\hbar}{2}\hat{P} \right],
	\end{equation*}
	where we used $[\hat{G},\hat{P}] = i\hbar \hat{P}$, the previous equation can be rewritten in a more "symmetric" form
	\begin{equation}\label{NewTimeEvolution}
	\begin{split}
	d|\Phi_t \rangle = - \frac{i}{\hbar}\hat{H}'|\Phi_t\rangle dt  + \sum_{k = a,b} \left(  \frac{i \sigma_k(t)}{\hbar l_t} \hat{F}_k|\Phi_t\rangle dW_t^k -  \frac{ \sigma_k(t)^2}{2\hbar^2 l_t^2} \hat{F}_k^2 |\Phi_t\rangle dt  \right)
	\end{split}
	\end{equation}
	with
	\begin{equation*}
	\begin{split}
	\hat{H}' &:= \hat{K} - \frac{\mu_1(t)}{l_t} \hat{P} - \mu_2(t)\hat{G}  +  \mbox{sign}(a_t - b_t)\frac{\sigma_a^2(t) - \sigma_b^2(t)}{2l_t^2} \hat{P}, \\
	\hat{F}_k &:= \frac{\hat{P}}{2} + \varepsilon_k\mbox{sign}(a_t - b_t)\hat{G} .
	\end{split}
	\end{equation*}
	This is the SDE describing the time-evolution of \eqref{VectorMapping} on the Hilbert space $L_2(\left[ -\frac{1}{2}, \frac{1}{2}\right]) \oplus L_2(\left[ -\frac{1}{2}, \frac{1}{2}\right]^c) = L_2(\Rea)$. However we are interested only in the time evolution taking place in $L_2(\left[ -\frac{1}{2}, \frac{1}{2}\right])$, i.e. on the evolution of the vector $|\varphi_t\rangle := \hat{W}|_{L_2[-1/2,1/2]} |\psi_t\rangle$. This can be derived from \eqref{NewTimeEvolution} simply by restricting all the operators to this subspace.
	The resulting equation is formally identical to \eqref{NewTimeEvolution} and is solved imposing the boundary conditions dictated by the Hamiltonian domain $D(\hat{H})$.
		
	Because of that, the domain of the Hamiltonian $\hat{H}$ is relevant. Consider the	following domain
	\begin{equation*}
		\mathcal{D}(\hat{H}) = \bigg\{ \psi \in H^2([a_t,b_t]) \bigg| \psi(a_t) = e^{i\theta}\psi(b_t), \frac{d\psi}{dx}(a_t) = e^{i\theta} \frac{d\psi}{dx}(b_t) \bigg\}
	\end{equation*}
	with $\theta \in [0,2\pi]$, namely the Sobolev space of square integrable functions having square integrable second derivative fulfilling quasi-periodic boundary conditions at the end points. $\theta$ is a parameter which cannot be fixed by mathematical considerations but is the physics of the problem that dictates the correct value \cite{reed1975ii}. For example, when $\theta = 0$ we have the periodic boundary conditions, used to simulate infinite periodic structures. On this domain the Kinetic term \eqref{Kinetic_term} of the Hamiltonian is self-adjoint. In addition, the term $\hat{P}^2$ appearing in the kinetic term of the particle Hamiltonian is the square of (self-adjoint) momentum operator $\hat{P}$ having domain
	\begin{equation*}
		D(\hat{P}) = \{ \psi,\psi' \in L_2\left(\left[ a_t, b_t\right]\right)  | \psi\left(a_t\right) = e^{i\theta}\psi\left(b_t\right) \},
	\end{equation*} 
	with $\theta$ the same used in defining the domain of $\hat{H}$ (in the sense that $\hat{P}^2$ and $\hat{P}$ have the same generalized eigenvector) \cite{bonneau2001self}. Using this momentum operator in \eqref{NewTimeEvolution} and passing from the \Ito to the Stratonovich formalism, the unitarity of the evolution map induced by \eqref{NewTimeEvolution} becomes explicit. In particular, the Stratonovich equation is similar to \eqref{NewTimeEvolution} except that the last term is missing and the \Ito differential is replaced by the Stratonovic differential. 
	The unitarity of the time-evolution is not a surprising fact, since the operator $\hat{W}|_{L_2[-1/2,1/2]} $ and the time evolution operator in $L_2([a_t,b_t])$ are unitary by construction.
	
	Let us now consider a different domain for $\hat{H}$, namely
	 \begin{equation*}
	 	\mathcal{D}(\hat{H}) = \{ \psi \in H^2\left([a_t,b_t]\right) | \psi(a_t) = \psi(b_t) = 0 \},
	 \end{equation*}
	 where the Dirichlet boundary conditions are used. With these boundary condition the kinetic term \eqref{Kinetic_term} is self-adjoint but the momentum operator is not. Hence in this case $\hat{W}|_{L_2[-1/2,1/2]} $ 
	is not unitary anymore, therefore we cannot conclude that the time-evolution operator associated to \eqref{NewTimeEvolution}, restricted on $L_2(\left[ -\frac{1}{2}, \frac{1}{2}\right])$, is unitary by using the same argument as before. However, it turns out that also in this case the time-evolution is indeed unitary. This can be proved by deriving the description of a particle in box with Dirichlet boundary conditions from that of a quantum particle moving in a finite potential well of height $v_0$, taking then the limit $v_0 \rightarrow \infty$. A proof is presented in the appendix. 
 
 	To conclude, suppose that the states $|\varphi_t^i\rangle$ evolve with \eqref{NewTimeEvolution} for any $i = 1, 2, \cdots$. The map for the density matrix $\hat{\rho}_t:= \sum_i p_i|\varphi_t^i \rangle\langle \varphi_t^i|$ can be easily derived. It turns out that it is a linear map of the form
	\begin{equation}\label{StochasticMap}
	d \hat{\rho}_t = \mathcal{L}_t(\hat{\rho}_t) dt + d\mathcal{W}_t(\hat{\rho}_t)
	\end{equation}
	where $\mathcal{L}_t(\cdot)$ takes the form of an ordinary Lindbladian but with stochastic and time-dependent coefficients
	\begin{equation*}
	\mathcal{L}_t(\cdot) =  - \frac{i}{\hbar}\left[ \hat{H}', \cdot \right] + \sum_{k = a,b} \frac{\sigma_k(t)^2}{\hbar^2 l_t^2 } \left( \hat{F}_k \cdot \hat{F}_k - \frac{1}{2} \bigg\{\hat{F}_k^2, \cdot \bigg\} \right),
	\end{equation*}
	while $d\mathcal{W}_t (\cdot)$ is a stochastic term taking the form
	\begin{equation*}
	d\mathcal{W}_t(\cdot) = \sum_{k = a,b} \frac{i \sigma_k(t)}{\hbar l_t} \bigg[\hat{F}_k , \cdot \bigg]dW_t^k.
	\end{equation*}
	Written in the Stratonovich formalism, the generator $\mathcal{L}_t(\cdot) dt + d\mathcal{W}_t(\cdot)$ can be easily recognized as a (stochastic) Liouvillian superoperator \cite{breuer2002theory}.
	
	\subsection{Physical interpretation}\label{2.4}
	
	Let us conclude our analysis discussing the physical interpretation of the result here obtained. Consider for a moment the average density matrix $\Ex[\hat{\rho}_t]$ and its time-evolution. By the property of the \Ito integral, if we take the expectation value of \eqref{StochasticMap} the stochastic term vanishes, $\Ex[ d\mathcal{W}_t(\hat{\rho}_t) ] = 0$. However for the averaged Lindbladian one has $\Ex[ \mathcal{L}_t(\hat{\rho}_t) ]\neq\mathcal{L}_t(\Ex[\hat{\rho}_t])$ due to the presence of the stochastic coefficients. Hence one cannot write down a closed equation for  $\Ex[\hat{\rho}_t]$. This however does not pose a problem as also observable quantities are stochastic and therefore physical quantities take the form $\Ex[\langle \varphi_t | \hat{A}'| \varphi_t \rangle]$, with $\hat{A}'$ suitably defined, as we will now discuss. 
	
	Taken an observable $A$, represented by a self-adjoint operator $\hat{A}$ on $L_2([a_t,b_t])$, one can map it on $L_2(\left[ -\frac{1}{2}, \frac{1}{2}\right])$ through $\hat{W}$. In order to do this, one first extends $\hat{A}$ to the whole Hilbert space $L_2([a_t,b_t])\oplus L_2([a_t,b_t]^c)$ obtaining $\hat{A}_{ext} = \hat{A}\oplus_{m_t,l_t}\hat{\Nullo}$, as done with $\hat{H}_{ext}$. Recalling that $\psi \in L_2([a_t,b_t])$ extended to $L_2([a_t,b_t])\oplus L_2([a_t,b_t]^c)$ is $\Psi = \psi + \phi$ as explained in section \ref{2.1}, by unitarity of $\hat{W}$, we can write that
	\begin{equation*}
	\begin{split}
	\langle \psi_t | \hat{A} |\psi_t \rangle &= \langle \Psi_t | \hat{A}_{ext} |\Psi_t \rangle |_{L_2[a_t,b_t]}
	= \langle \Psi_t |\hat{W}^{\dagger}\hat{W} \hat{A}_{ext} \hat{W}^{\dagger}\hat{W}|\Psi_t\rangle|_{L_2[a_t,b_t]}  \\
	&=\langle \Phi_t | \hat{A}' | \Phi_t \rangle|_{L_2[-1/2,1/2]} = \langle \varphi_t | \hat{A}'| \varphi_t \rangle,
	\end{split}
	\end{equation*}
	where $\hat{A}':= \hat{W}(m_t,l_t) \hat{A}_{ext} \hat{W}^{\dagger}(m_t,l_t)$ (we used the same symbol $\hat{A}'$ when considering the restriction to $L_2(\left[ -\frac{1}{2}, \frac{1}{2}\right])$) and again we set $| \varphi_t \rangle = \hat{W}|_{L_2[-1/2,1/2]}|\psi_t\rangle$. It is not difficult to see that $\hat{A}'$, the operator representing the observable $A$ on $L_2(\left[ -\frac{1}{2}, \frac{1}{2}\right])$, is stochastic. As consequence of that, the expectation value of the observable $A$ is
	\begin{equation*}
	\Ex[A]  = \int_\Omega \langle \varphi_t(\omega) | \hat{A}'(\omega) | \varphi_t(\omega) \rangle P(d\omega) 
	= \int_\Omega \sum_{a' \in \sigma(\hat{A}')} a'(\omega) | \langle a'(\omega) | \varphi_t(\omega) \rangle|^2 P(d\omega). 
	\end{equation*}
	where we assumed $\hat{A}'$ to be a compact operator, for simplicity, and we used its spectral decomposition. Since the vector $|\varphi_t\rangle$ remains normalized under the time-evolution \eqref{NewTimeEvolution}, the quantity $P(A = a'| \omega) = | \langle a'(\omega) | \varphi_t(\omega) \rangle|^2 $ can be interpreted as a conditional probability. This suggests the correct interpretation of $|\varphi_t(\omega)\rangle$: it is the state of the quantum particle given a specific realization of the boundary. At the level of the density matrix, the above considerations implies that the expectation value is given by
	\begin{equation*}
	\Ex[A] = \int_\Omega \Tr{ \hat{\rho}_t\hat{A}'}P(d\omega)
	\end{equation*}
	where $\hat{\rho}_t$ fulfils \eqref{StochasticMap}. This means that $\hat{\rho}_t$, not its averaged version $\Ex[\hat{\rho}_t]$, represents the state of our quantum system, as anticipated in the beginning. Summarizing, because of the stochasticity of the observables, the time-evolution of the state here derived cannot be written in Lindblad form.
	%
	%
	
	\subsection{Example: Quantum particle trapped between two ideal mirrors.}\label{Examples}
	
	As an application, we consider the case of a quantum particle bouncing back and forth between two ideal mirrors at temperature $T$. With the word \textquotedblleft ideal\textquotedblright $\mspace{1mu}$we mean that the particle can never escape from the region between the two mirrors, which mathematically imply the Dirichlet boundary conditions. However, the equations remain valid also for the case of quasi-periodic boundary conditions. Obviously, physical mirrors cannot  penetrate each other and this feature has to be taken into account in order to describe realistic situations. A very simple and physically reasonable mathematical way to describe the motion of the mirrors, which is able to reproduce this basic feature, is provided by the Dyson brownian motion. In the first example we will consider this case, which can be seen as a prototype for physically meaningful mirrors.  In the remaining two examples we consider the Langevin dynamics in the ordinary and over-damped regime, to describe the dynamics of an object in an environment at temperature $T$. 
	
	\paragraph{Dyson brownian motion of the mirrors.}
	As a first example, we consider the case where the boundary motion is described by two Dyson brownian motions \cite{dyson1962brownian}:
	\begin{equation}\label{Dyson}
	\begin{cases}
		da_t = \frac{\beta}{a_t - b_t}dt + \sigma_a dW_t^{a} \\
		db_t = \frac{\beta}{b_t - a_t}dt + \sigma_a dW_t^{b}
	\end{cases}
	\end{equation}
	where $\beta \in \Rea$ is an arbitrary parameter. The Dyson brownian motion is interesting because it has the property that the two diffusion processes are non-colliding. This can be seen directly from the drift term: as the two boundaries try to collide, the drift term tend to push them in opposite directions. This means that if we start with $a(0) < b(0)$, then this remains true for all later times. In equation \eqref{NewTimeEvolution} this means that $\mbox{sign}(a_t-b_t) = -1$. More important, the situation with $l_t=0$ never happens, which means that \eqref{NewTimeEvolution} is valid for all times. The physical interpretation of this motion for the mirrors is quite natural. The drift term models an electrostatic repulsive potential between the two boundaries (hence $\beta$ plays the role of an effective coupling constant). This can be seen as a very simple way to take into account that mirrors are made of matter which interacts electromagnetically.
	
	Using \eqref{Dyson} we obtain the following quantities
	\begin{equation*}
	\begin{split}
	\mu_1(t) &= 0 \\
	\mu_2(t)&= \frac{4\beta - \sigma_a^2(t)-\sigma_b^2(t)}{2l_t^2}
	\end{split}
	\end{equation*}
	The time evolution of the particle with Hamiltonian $\hat{H}$ on $L_2[a_t,b_t]$ is given by equation \eqref{NewTimeEvolution} restricted to $L_2(\left[ -\frac{1}{2}, \frac{1}{2}\right])$ with
	\begin{equation}
	\begin{split}
	\hat{H}' &:= \hat{K} + \frac{4\beta - \sigma_a^2(t)-\sigma_b^2(t)}{2l_t^2}\hat{G}  - \frac{\sigma_a^2(t) - \sigma_b^2(t)}{2l_t^2} \hat{P}, \\
	\hat{F}_k &:= \frac{\hat{P}}{2} - \varepsilon_k\hat{G},
	\end{split}
	\end{equation}
	where $\hat{K} = \hat{W}\hat{H} \hat{W}^{\dagger}$. For a free particle, $\hat{K} = \hat{P}^2/2ml_t^2$, where we used $\hat{W}\hat{P} \hat{W}^{\dagger} = \hat{P}/l_t$. This example shows that a very simple way to avoid the unpleasant situation $l_t=0$, namely that the two boundaries penetrate each other (i.e. $a_t < b_t$ for all times), is to add a term $2\beta \hat{G}/ l_t^2$ in the Hamiltonian $\hat{H}'$. Note that $\beta$ can be chosen arbitrarily small. We will use this observation in the following two examples.
	
	\paragraph{Langevin dynamics of the mirrors.} The Langevin dynamics of $a_t$ is described by the following equations\footnote{Typically, the Langevin dynamics is described in the Stratonovich formalism \cite{van1992stochastic}. However, since the noise is white and additive, the \Ito and the Stratonovich descriptions coincide.}:
	\begin{equation}\label{langevin}
	\begin{cases}
	da_t = v_t^{a}dt \\
	mdv_t^{a} = -\gamma v_t^{a}dt + DdW_t^a
	\end{cases}
	\end{equation}
	with $D = \sqrt{2k_bT\gamma}$, where $k_b$ is the Boltzman constant and $\gamma$ is the damping constant. The same equation holds for $b_t$, assuming that the two mirrors have the same mass $m$. Solving the SDE, one obtains the following equations for the extrema of the interval
	\begin{equation*}
	da_t = (v^a(0) + DW_t^a)e^{-\frac{\gamma}{m}t}dt, \mspace{50mu}
	db_t = (v^b(0) + DW_t^b)e^{-\frac{\gamma}{m}t}dt,
	\end{equation*}
	where $v^a(0)$ and $v^b(0)$ are the initial velocities of $a_t$ and $b_t$, respectively. We assume that the two mirrors remain sufficiently far apart for at least some time $t_{max}$ sufficiently big, i.e. $a_t < b_t$ from $0 \leqslant  t < t_{max}$. From the equations above, one can see that
	\begin{equation*}
	\begin{split}
	\mu_1(t) &= \frac{v^a(0)+v^b(0)+D(W_t^a + W_t^b)}{2}e^{-\frac{\gamma}{m}t} \\
	\mu_2(t)&= -\frac{v^a(0)-v^b(0)+D(W_t^a - W_t^b)}{l_t}e^{-\frac{\gamma}{m}t}
	\end{split}
	\end{equation*}
	where we used $\mbox{sign}(a_t-b_t)=-1$ since we assumed $a(0)<b(0)$. The time evolution of the particle given by equation  \eqref{NewTimeEvolution} and restricted to $L_2(\left[ -\frac{1}{2}, \frac{1}{2}\right])$ reduces to
	\begin{equation}
	\begin{split}
	d|\varphi_t \rangle = - \frac{i}{\hbar}\bigg[ &\hat{K} - \frac{v^a(0) + v^b(0)+D(W^a_t + W^b_t)}{2l_t}e^{-\frac{\gamma}{m}t}\hat{P} \\
	&+  \frac{v^a(0) - v^b(0)+D(W^a_t - W^b_t)}{l_t}e^{-\frac{\gamma}{m}t}\hat{G}\bigg]|\varphi_t\rangle dt.
	\end{split}
	\end{equation}
	The first term in the RHS gives the energy of the quantum particle, while the last two terms are the corrections due to the time-dependency of the original Hilbert space. Note that, even if the extrema of the interval evolve randomly, we do not get a stochastic differential equation for the state-vector. Nevertheless this time-evolution is random since it depends on two random variables.
	
	The equations derived here depend on the fact that there exist some $t_{max}$ such that $a_t < b_t$ for all $t < t_{max}$. To avoid using this condition, one can always introduce an electrostatic potential among the boundaries, as in the example of the Dyson brownian motion seen before. In general the effect of adding this kind of drift term is the following:
	\begin{equation*}
		\mu_2(t) \rightarrow \mu_2(t) + \frac{2\beta}{l_t^2},
	\end{equation*}
	for some $\beta \in \Rea$, while $\mu_1(t)$ does not change. As already observed, this means adding a term $2\beta \hat{G}/ l_t^2$ to the effective Hamiltonian. Hence for $\beta$ sufficiently small, the description given above is a very good approximation of the more realistic situation considered here. Note that $\beta << 1$ is a quite natural assumption since the two mirrors experience an electrostatic repulsion only when they are very close to each other (this means that the coupling constant is very small).

	\paragraph{Over-damped Langevin dynamics of the mirrors.} The over-damped Langevin dynamics is realized when the inertia of the body is negligible compared to the random force. In our case this corresponds to the case of "light mirrors". In this regime, the average acceleration of the two mirrors can be considered equal to zero. Thus equations \eqref{langevin} for the time evolution of the mirrors' positions are well approximated by 
	\begin{equation*}
	da_t = \frac{D}{\gamma} dW_t^a   \mspace{50mu} db_t = \frac{D}{\gamma} dW_t^b.
	\end{equation*}
	From these SDEs, one immediately derives that $\mu_1 = 0$ and
	\begin{equation*}
	\mu_2 = - \frac{D^2}{l_t^2\gamma^2}.
	\end{equation*}
	Moreover, $\sigma_a(t) = \sigma_b(t) = D/\gamma$. Also in this case we work under the condition that there exist some $t_{max}$ such that $a_t < b_t$ for all $t < t_{max}$. As before, to avoid the use of this condition, one can add a suitable electrostatic repulsive potential. In this regime, the time evolution of the particle as given by \eqref{NewTimeEvolution} reduces to
	\begin{equation}\label{predi}
	d|\varphi_t \rangle = - \frac{i}{\hbar}\left(\hat{K} + \frac{D^2}{l_t^2\gamma^2}\hat{G}\right)|\varphi_t\rangle dt  + \sum_{k = a,b} \left( \frac{i D}{\hbar l_t\gamma} \hat{F}_k|\varphi_t\rangle dW_t^k -  \frac{ D^2}{2\hbar^2 l_t^2\gamma^2} \hat{F}_k^2 |\varphi_t\rangle dt  \right),
	\end{equation}
	with
	\begin{equation*}
	\hat{F}_k = \frac{\hat{P}}{2} - \varepsilon_k\hat{G}.
	\end{equation*}
	Particularly interesting is the case of the Harmonic oscillator, i.e.
	\begin{equation*}
		\hat{H} = \frac{\hat{P}^2}{2} +  \frac{\omega^2}{2}\hat{X}^2
	\end{equation*}
	(with $m=1$).
	Using this Hamiltonian in \eqref{predi} and rewriting the equation in terms of the creation and annihilation operators on $L_2(\left[ -\frac{1}{2}, \frac{1}{2}\right])$\footnote{To do this operation in the proper manner one should rewrite \eqref{predi} in terms of annihilation and creation operators before to restrict the Hilbert space to $L_2(\left[ -\frac{1}{2}, \frac{1}{2}\right])$. }, namely
	\begin{equation*}
		\hat{a}' := \frac{1}{\sqrt{2}}\left( \sqrt{\frac{\omega}{\hbar}}\hat{P}' + i \frac{1}{\sqrt{\hbar \omega}} \hat{X}' \right)  \mspace{30mu} \hat{a}^\dagger := \frac{1}{\sqrt{2}}\left( \sqrt{\frac{\omega}{\hbar}}\hat{P}' - i \frac{1}{\sqrt{\hbar \omega}} \hat{X}' \right),
	\end{equation*}
	with $\hat{P}' = \hat{W}\hat{P} \hat{W}^{\dagger}$ and $\hat{X}' = \hat{W}\hat{X} \hat{W}^{\dagger}$, we can easily see that the second term in the drift of \eqref{predi} is a squeezing-like term \cite{lvovsky2015squeezed}. The squeezing parameter turns out to be $\zeta = \zeta ^* = 2D^2/\gamma^2l_t^2$, which is stochastic and inversely proportional to the distance between the two  mirrors, as one should expect from physical considerations. Indeed we expect that when the mirrors get closer enough, the particle's wave function gets more localized in the position basis. Note that this squeezing-like term has a purely stochastic origin: it would disappear if the stochasticity goes to zero (i.e. $\sigma_i \rightarrow 0$, with $i=a,b$).\newline
	When the system is a photon (which, when confined in a cavity, can be effectively described by an harmonic oscillator), the additional squeezing-like term we get in the drift is compatible with what considered in \cite{law1994effective}, where the deterministic case is studied by using field theoretic techniques. Note that we get only a single-mode description, since we are considering a single particle Hamiltonian. The multi-mode case can be also obtained in this framework by using many-particle Hamiltonians with suitable two-particle interactions.\newline
	
	Note that in all cases of the examples, no Lindblad-type dynamics is produced by the vibrating mirrors. In fact, as it should be clear from the derivation, we do not model any interaction between the mirrors and particle. When this is not possible a different approach, based on quantum Brownian motion, should be used \cite{law1995interaction,giovannetti2001phase}.
	
	\section{Time-evolution in Hilbert spaces with stochastic measure}\label{3}
	
	Let us now consider a quantum system described at each instant of time $t \in \Rea$ by a vector belonging to an Hilbert space $L_2(\mathcal{M},m_t(x)dx)$. The stochastic measure $\mu(dx) = m_t(x)dx$ is assumed to be positive, in order to have a positive-definite scalar product. To define any kind of time-evolution, the first issue that we have to face is again the definition of the time-derivative: vectors at different times are defined on different Hilbert spaces. This can be realized again by mapping $L_2(\mathcal{M},m_t(x)dx)$ to the time-independent Hilbert space $L_2(\mathcal{M},dx)$ by defining a suitable operator $\hat{h}_t$ which takes into account the effects of the time-dependent stochastic part of the measure $m_t(x)dx$. 
	
	\subsection{Mapping on a time-independent Hilbert space}\label{hhdagasection}
	
	Let us assume that, as function of time, $m_t(x)$ fulfills the following SDE
	\begin{equation}\label{measurevolution}
	d m_t(x) =  \mu_t (x)dt + \sigma_t(x)dW_t,
	\end{equation}
	where $\mu_t(x)$ and $\sigma_t(x)$ are the drift and variance functions. This SDE is defined on a a probability space $(\Omega,\E,P)$ and $W_t$ is a standard Wiener process adapted to a given filtration $\{\mathcal{F}_t\}_{t \in \Rea^+}$. On $L_2(\mathcal{M},m_t(x)dx)$ the scalar product is defined as
	\begin{equation*}
	\langle \psi | \phi \rangle_t := \int_{\mathcal{M}} \psi^*(x)\phi(x)m_t(x)dx,
	\end{equation*}
	for any $\psi$, $\phi \in L_2(\mathcal{M},m_t(x)dx)$. In order to have a well defined scalar product at each time $t$, we assume that the function $m_t(x)$ and its time evolution are such that 
	\begin{equation}\label{boundedop}
	0 < H \leqslant m_t(x) \leqslant K < \infty \mspace{50mu} \forall t \in \Rea, \forall x \in \mathcal{M},
	\end{equation}
	where $H,K \in \Rea^+$. We will comment on this condition later. Consider now the usual Hilbert space $L_2(\mathcal{M},dx)$, whose scalar product is
	\begin{equation*}
	\langle \tilde{\psi} | \tilde{\phi} \rangle_2 := \int_\mathcal{M}  \tilde{\psi}^*(x)\tilde{\phi}(x)dx, 
	\end{equation*}
	for any $\tilde{\psi}$, $\tilde{\phi} \in L_2(\mathcal{M},dx)$. Note that if $\psi \in L_2(\mathcal{M},m_t(x)dx)$ then $\psi \in L_2(\mathcal{M},dx)$ and vice versa, because of \eqref{boundedop}. Following \cite{mostafazadeh2004time}, let us define the positive operator
	\begin{equation}\label{measureoperator}
	\hat{m}_t \tilde{\psi}(x) = m_t(x)\tilde{\psi}(x),
	\end{equation}
	which acts on $L_2(\mathcal{M},dx)$. From \eqref{boundedop} we conclude that it is a bounded operator and, being a positive operator, it admits a unique positive square root \cite{moretti2013spectral}. Hence on $L_2(\mathcal{M},dx)$ there exists a positive operator $\hat{h}_t$ such that $\hat{m}_t = \hat{h}_t^2$. Since $\hat{h}_t$ is positive then $\hat{h}_t = \hat{h}_t^{\dagger_2}$, where $^{\dagger_2}$ is the adjoint according to the scalar product of $L_2(\mathcal{M},dx)$. Comparing $\langle \cdot| \cdot \rangle_t$ and $\langle \cdot | \cdot \rangle_2$, for any $\psi,\phi \in L_2(\mathcal{M},m_t(x)dx)$ we can write
	\begin{equation}\label{turcofondamentale}
	\begin{split}
	\langle \psi | \phi \rangle_t = \langle \psi | \hat{m}_t \phi \rangle_2 = \langle \hat{h}_t \psi | \hat{h}_t \phi \rangle_2.
	\end{split}
	\end{equation}
	Hence for each $\psi \in L_2(\mathcal{M},m_t(x)dx)$ we can associate a vector $\tilde{\psi} = \hat{h}_t \psi \in L_2(\mathcal{M},dx)$. This means that $\hat{h}_t$ can be considered also as an operator between the two Hilbert spaces considered here, i.e. $\hat{h}_t : L_2(\mathcal{M},m_t(x)dx) \rightarrow L_2(\mathcal{M},dx)$ providing a mapping between the time-dependent Hilbert space to a time-independent one at any time. When considered in this way, from the general definition of adjointness for operators between two different Hilbert spaces \footnote{Given a linear operator $\hat{G}:\Hi_1 \rightarrow \Hi_2$, its \emph{adjoint} is the operator $\hat{G}^{\dagger}:\Hi_2 \rightarrow \Hi_1$ such that $	\langle u | \hat{G}v  \rangle_2 = \langle \hat{G}^\dagger u | v \rangle_1$ where $u \in \Hi_2$ and $v \in \Hi_1$\cite{moretti2013spectral}.}, we have that $\hat{h}_t \neq \hat{h}_t^{\dagger}$, where $\hat{h}_t^\dagger : L_2(\mathcal{M},dx) \rightarrow L_2(\mathcal{M}, m_t(x)dx)$. This observation will be important in the next section since it will allow to define the Hilbert space which describes our quantum particle at any time.
	
	\subsection{The Hilbert of the quantum system}\label{reproducing-kernel}
	
	In order to describe the time-evolution of our quantum system, we need to specify the Hilbert space used to describe our quantum system at any time. First we note that $\hat{h}_t$ allows to define a scalar product between vectors belonging to two different Hilbert spaces (which are nothing but the transition amplitudes of the particle). More precisely, taken $\psi_t \in L_2(\mathcal{M},m_t(x)dx)$ and $\phi_s \in L_2(\mathcal{M},m_s(x)dx)$, we may set
	\begin{equation}\label{elemtaryscalarp}
	\langle \psi_t | \phi_s \rangle := \langle \hat{h}_t \psi_t | \hat{h}_s \phi_s \rangle_2 = \langle \psi_t | \hat{h}_t^{\dagger}\hat{h}_s \phi_s \rangle_t.
	\end{equation}
	Note that for $t=s$ this definition reduces to the usual scalar product on $L_2(\mathcal{M},m_t(x)dx)$, implying that $\hat{h}_t^{\dagger}\hat{h}_t = \Id$. Hence $\hat{h}_t$ is a unitary mapping between $L_2(\mathcal{M},m_t(x)dx)$ and $L_2(\mathcal{M},dx)$. A proof that \eqref{elemtaryscalarp} is an inner product is given in appendix. At this point, we are ready to define the Hilbert space on which we can describe the particle at any time $t$. We will label such a space by $\Hi_K$. The minimal requirement for $\Hi_K$ to describe the quantum system at any time is to contain all vectors $ \psi_t \in L_2(\mathcal{M},m_t(x)dx)$ for any $t \in \Rea$. Hence we may define $\Hi_K$ as the linear space
	\begin{equation*}
		\Hi_K := \underset{{t \in \Rea}}{\mbox{span }} \{ L_2(\mathcal{M},m_t(x)dx) \}.
	\end{equation*}	
	When equipped with the inner product induced by \eqref{elemtaryscalarp} $\Hi_K$ is an inner product space. As such, it can always be completed in order to be an Hilbert space \cite{moretti2013spectral}. In what follows $\Hi_K$ denotes also its completion. Note that $\Hi_K$ is separable, since at any time $t$ any $\psi_t$ can be mapped to $L_2(\mathcal{M},dx)$, which is separable. Our quantum system at each time $t \in \Rea$ is described by some $\psi_t \in L_2(\mathcal{M},m_t(x)dx)$, which is also a vector in $\Hi_K$. Hence, the dynamics of our quantum system is always described by a vector of $\Hi_K$ and for this reason is chosen as the Hilbert space associated to our particle, at any time. This choice enables us to define a time derivative of the state, since now $\psi_{t+\epsilon}$ and $\psi_t$ belong to the same Hilbert space and can be both mapped on $L_2(\mathcal{M},dx)$.
	
	\subsection{Derivation of the time-evolution}\label{derivationtimeevolution}
	
	Given the Hilbert space $\Hi_K$ a unitary evolution can be introduced. Let $\hat{H}$ be the Hamiltonian of our quantum system at time $t$. It is an operator on $L_2(\mathcal{M},m_t(x)dx)$, hence it has a time-dependent stochastic domain. More precisely,  we may chose
	\begin{equation*}
	\mathcal{D}(\hat{H}) = \{\psi \in L_2(\mathcal{M},m_t(x)dx)| \hat{H}\psi \in L_2(\mathcal{M},m_t(x)dx)\},
	\end{equation*}
	and we also assume $\hat{H}$ to be self-adjoint on this domain. As done in section \ref{2.1}, one can extend $\hat{H}$ to the whole $\Hi_K$ by setting $\hat{H}_{ext} := \hat{H} \oplus_t \hat{\Nullo}$, where $\hat{\Nullo}$ is the null operator on $L_2(\mathcal{M},m_t(x)dx)^{\perp}$\footnote{The symbol means $^\perp$ orthogonal subspace with respect to $\Hi_K$.} (in particular we chose $|\phi\rangle = 0$). Given $|\psi_t\rangle \in L_2(\mathcal{M},m_t(x)dx)$ one extends it on $\Hi_K$ simply setting $|\Psi_t\rangle := |\psi_t\rangle + | \phi \rangle$, where $|\phi\rangle$ is an arbitrary vector on $L_2(\mathcal{M},m_t(x)dx)^{\perp}$. At this point, the unitary time-evolution is given by
	\begin{equation}\label{unitarydim}
	d|\Psi_t\rangle = - \frac{i}{\hbar}\hat{H}_{ext} |\Psi_t\rangle dt.
	\end{equation}
	Again $|\Psi_t\rangle$ is a stochastic process of bounded variation and the equation is understood in the sense of stochastic integrals. Using the scalar product \eqref{elemtaryscalarp}, one can easily check that $d\| \Psi_t \|^2 = 0$ under this time evolution and $|\Psi_t\rangle \in L_2(\mathcal{M},m_t(x)dx)$ for any $t$. Despite equation \eqref{unitarydim} displays explicitly the unitarity of the time-evolution, to do explicit calculations one has to represent it in some particular basis. Below we present two possible representations. The example done in section \eqref{example} will clarify their physical relevance.
	
	\paragraph{Representation on the position basis of $L_2(\mathcal{M},dx)$.} 
	Recalling that any $|\Psi_t\rangle \in L_2(\mathcal{M},m_t(x)dx) \subset \Hi_K$ also belongs to $L_2(\mathcal{M},dx)$, we can use the generalized eigenvectors of the position operator $\{|x\rangle\}$ on $L_2(\mathcal{M},dx)$. Setting $\Phi_t(x) = \langle x | \Psi_t \rangle_2$ and using \eqref{elemtaryscalarp} and \eqref{turcofondamentale}, we can write that
	\begin{equation*}
	\Phi_t(x) = \langle x | \Psi_t \rangle_2 = \langle x | \hat{m}_t^{-1}\hat{m}_t\Psi_t \rangle_2 = \langle x| \hat{m}_t^{-1} \Psi_t \rangle
	\end{equation*}
	Thus, the time evolution is
	\begin{equation*}
	\begin{split}
	d\Phi_t(x) &= d \langle x| \Psi_t \rangle_2 = d \langle x| \hat{m}_t^{-1} \Psi_t \rangle \\
	&= \langle x | d\hat{m}_t^{-1}  \Psi_t \rangle + \langle x | \hat{m}_t^{-1} d \Psi_t \rangle \\
	&= \langle x | \frac{d\hat{m}_t^{-1}}{\hat{m}_t^{-1}} \hat{m}_t^{-1}  \Psi_t \rangle + \langle x |  d \Psi_t \rangle_2 \\
	&= \langle x | d\hat{m}_t^{-1} \hat{m}_t \Psi_t \rangle_2 + \langle x |  d \Psi_t \rangle_2.
	\end{split}
	\end{equation*}
	Expressing all the operators in the position basis the above equation can be rewritten as 
	\begin{equation}\label{EQrep1}
	d\Phi_t(x) = -\frac{i}{\hbar}\hat{H}_{ext} \Phi_t(x) dt + d\hat{m}_t^{-1}\hat{m}_t\Phi_t(x).
	\end{equation}
	Note that in this equation, the stochastic differential must be computed using the \Ito formula in equation \eqref{measurevolution}.
	From the general theory of stochastic processes \cite{klebaner2005introduction}, the  operator in the correction term, $d\hat{m}_t^{-1}\hat{m}_t$, is the stochastic differential of the stochastic logarithm of $\hat{m}_t^{-1}$, i.e. $d\Slog(\hat{m}_t^{-1})$ (see appendix). Note that, despite the presence of an hermitian term, the time-evolution is still unitary: the additional term simply reflects the change of the measure $m_t(x)$. To see it explicitly, one may compute the equation for $\Phi_t(x)^*$. Since
	\begin{equation*}
	\Phi_t(x)^* =  [\langle x | \Psi_t \rangle_2]^* = [\langle x| \hat{m}_t^{-1} \Psi_t \rangle]^* = \langle \Psi_t \hat{m}_t^{-1}| x \rangle,
	\end{equation*} 
	and $\hat{m}_t^{\dagger} = \hat{m}_t^{-1}$ in $\Hi_K$ (see appendix), we have that $\Phi_t(x)^* = \langle \Psi_t| \hat{m}_t x \rangle$. Hence the evolution of $\Phi_t(x)^*$ is given by the equation \eqref{EQrep1} but with the operator in the last term replaced by $d\hat{m}_t\hat{m}_t^{-1} = d \Slog(\hat{m}_t)$. Note that this result can be obtained by taking the dagger in $\Hi_K$ of \eqref{EQrep1} directly. At this point, by direct computation
	\begin{equation*}
	\begin{split}
	d \| \Phi_t\|^2 &= d\int_{\mathcal{M}} \Phi_t(x)^*\Phi_t(x)dx \\
	&= \int_{\mathcal{M}} \left( d	\Phi_t(x)^*\Phi_t(x) + \Phi_t(x)^*d\Phi_t(x) + d\llbracket \Phi_t(x)^*,\Phi_t(x) \rrbracket \right) dx = \\
	&= \int_{\mathcal{M}} |\Phi_t(x)|^2 \left( d \Slog(\hat{m}_t^{-1}) + d\Slog(\hat{m}_t) + d\llbracket \hat{m}_t^{-1},\hat{m}_t \rrbracket \right) dx = 0
	\end{split}
	\end{equation*}
	where in the last step we used the properties of the stochastic logarithm proved in the appendix (see eq. \eqref{inverseslog}).
	
	\paragraph{Representation on the position basis of $\Hi_K$.} 
	
	The operator $\hat{h}_t$ allows to introduce a second representation. Given the generalized basis $\{|x\rangle\}$ in $L_2(\mathcal{M},dx)$, one can lift it to $\Hi_K$ by defining,
	\begin{equation}\label{newbase}
	|x,t\rangle := \hat{h}_t|x\rangle.
	\end{equation}
	This basis will be called position basis of $\Hi_K$. Note that the completeness relation for this basis holds when integrated on the whole $\mathcal{M}$ with respect to $dx$. At this point, one projects $|\Psi_t\rangle \in \Hi_K$ on this generalized basis obtaining the function $\chi_t(x) := \langle x,t | \Psi_t\rangle$. In this case one obtain
	\begin{equation*}
	\begin{split}
	d\chi_t(x) &= d \langle x,t | \Psi_t\rangle = d \langle x |\hat{h}_t^{\dagger} \Psi_t\rangle \\
	&= \langle x |d\hat{h}_t^{\dagger} \Psi_t\rangle + \langle x |\hat{h}_t^{\dagger} d\Psi_t\rangle \\
	&= \langle x |d\hat{h}_t^{\dagger}\hat{h}_t\hat{h}_t^{\dagger} \Psi_t\rangle + \langle x |\hat{h}_t^{\dagger}\left( -\frac{i}{\hbar} \hat{H}_{ext}\right)\Psi_t\rangle dt \\
	&= \langle x |d\hat{h}_t^{\dagger}\hat{h}_t\hat{h}_t^{\dagger} \Psi_t\rangle + \langle x |\hat{h}_t^{\dagger}\left( -\frac{i}{\hbar} \hat{H}_{ext}\right)\hat{h}_t\hat{h}_t^{\dagger}\Psi_t\rangle dt
	\end{split}
	\end{equation*}
	Calling $\hat{K}:=\hat{h}_t^{\dagger}\hat{H}_{ext}\hat{h}_t$ and expressing all the operators in the $|x\rangle$ basis, we obtain
	\begin{equation}\label{EQrep2}
	d\chi_t(x) = -\frac{i}{\hbar} \hat{K} \chi_t(x)dt + d\hat{h}_t^{\dagger}\hat{h}_t \chi_t(x)
	\end{equation}
	In this case, the stochastic differential must be computed using the \Ito formula for $\hat{h}_t^{\dagger} = \hat{m}_t^{-1/2}$. The same considerations on the norm conservation done in the previous paragraph apply, once one recognizes that $d\hat{h}_t^{\dagger}\hat{h}_t = d\hat{h}_t^{-1}\hat{h}_t = d\Slog(\hat{h}_t^{-1})$.
	Thus the time evolution is still unitary despite the presence of non-hermitian terms.\newline
	
	Before concluding let us comment on the condition \eqref{boundedop}. The careful reader may have noticed that this condition can be relaxed to
	\begin{equation}\label{relaxedcond}
		0 < H \leqslant m_t(x) < \infty \mspace{50mu} \forall t \in \Rea, \forall x \in \mathcal{M},
	\end{equation}
	with $H \in \Rea^+$. Under this relaxed condition, the scalar product \eqref{elemtaryscalarp} is still well defined and so is the Hilbert space $\Hi_K$. However, the difference with respect to the previous case is that now $\psi \in L_2(\mathcal{M},dx)$ does not imply $\psi \in L_2(\mathcal{M},m_t(x)dx)$. Hence the representation \eqref{EQrep1} still makes sense, but the representation \eqref{EQrep2} does not. Indeed, this last representation is defined by mapping (generalized) base vectors of $L_2(\mathcal{M},dx)$ on $L_2(\mathcal{M},m_t(x)dx)$, see \eqref{newbase}.

	\subsection{Example: unitary time-evolution of a quantum particle on a stochastic 3D manifold.}\label{example}
	
	Consider a 3-dimensional Riemannian manifold $\mathcal{M}$ with coordinates $\{x^i\}$ and metric tensor $g_{ij}(\mathbf{x})$. Assume that the metric tensor is a continuously differentiable function of the coordinates, i.e. $g_{ij}(\mathbf{x}) \in C^{\infty}(\mathcal{M})$ for any $i,j = 1,2,3$, and that the six independent components of $g_{ij}(\mathbf{x})$ are stochastic processes fulfilling for any $i \geqslant j$ (the remaining components are not independent due to the symmetry property of the metric tensor)
	\begin{equation}\label{stochasticmetric}
	dg_{ij}(\mathbf{x};t) = \alpha_{ij}(\mathbf{x};t)dt + \beta_{ij}(\mathbf{x};t)dW^{(ij)}_t. 
	\end{equation}
	Here $dW^{(ij)}_t$ are six independent standard Wiener processes\footnote{We used the parenthesis to indicate that $\beta_{ij}dW^{(ij)}_t$ is not $\sum_{ij}\beta_{ij}dW^{(ij)}_t$. We do not use the Einstein summation convention for the indexes of the Wiener process. However, in case of sum over repeated indexes like for example $k^{ij}\beta_{ij}dW^{(ij)}_t$, the Wiener processes are summed, i.e. $k^{ij}\beta_{ij}dW^{(ij)}_t = \sum_{ij}k^{ij}\beta_{ij}dW^{(ij)}_t$.  }. A stochastic metric implies a stochastic volume element $dV_t = \sqrt{ g_t(\mathbf{x})}d^3\mathbf{x}$ of the manifold $\mathcal{M}$, where $g_t(\mathbf{x}) = \det[g_{ij}(\mathbf{x};t)]$ as usual. Hence the Hilbert space associated to a quantum particle on $\mathcal{M}$, $L_2(\mathcal{M},dV_t)$, is time-dependent and stochastic. For the moment let us work assuming that the stochastic time evolution of $g_{ij}(\mathbf{x};t)$, is such that $\sqrt{g_t(\mathbf{x})}$ fulfills the condition \eqref{boundedop}. We can easily conclude that, for the problem considered here
	\begin{equation*}
	\hat{m}_t = \sqrt{g_t(\hat{\mathbf{X}})}, \mspace{30mu} \hat{h}_t = \sqrt[4]{g_t(\hat{\mathbf{X}})}, \mspace{30mu} \hat{h}_t^\dagger = \frac{1}{\sqrt[4]{g_t(\hat{\mathbf{X}})}}.
	\end{equation*}
	From now on we omit the arguments of operators above to keep the notation simple, when no confusion arises. The SDE for the volume element can be derived from \eqref{stochasticmetric}, obtaining
	\begin{equation}
	d \sqrt{\hat{g}} =  \frac{1}{2}\sqrt{\hat{g}} \bigg\{  \left[ \hat{g}^{ij} \hat{\alpha}_{ij} - \frac{(\hat{g}^{ij}\hat{\beta}_{ij})^2}{4}\right]dt +\hat{g}^{ij} \hat{\beta}_{ij}dW_t^{(ij)} \bigg\}.
	\end{equation}
	At this point we can write down two representations of the time evolution. 
	
	Let us consider the representation in $L_2(\mathcal{M},dx)$. Since $\hat{m}_t^{-1} = \hat{g}_t^{-1/2}$, equation \eqref{EQrep1} becomes
	\begin{equation}\label{Es1rep1}
	d\Phi_t(\mathbf{x}) = -\frac{i}{\hbar}\hat{H}_{ext}\Phi_t(\mathbf{x}) + d\Slog\left(\hat{g}_t^{-1/2}\right)\Phi_t(\mathbf{x})
	\end{equation}
	For the computation of the last term, one simply applies the \Ito formula, obtaining
	\begin{equation*}
	\begin{split}
	d\hat{g}_t^{-1/2} &= -\hat{g}_t^{-1/2}d\hat{g}_t^{-1/2} + \frac{1}{2}\hat{g}_t^{-3/2} d\llbracket \hat{g}_t^{-1/2},\hat{g}_t^{-1/2} \rrbracket \\
	&= - \frac{1}{2}\hat{g}_t^{-1/2} \left[ \left( \hat{g}^{ij}\hat{\alpha}_{ij} - \frac{3}{4} (\hat{g}^{ij}\hat{\beta}_{ij})^2 \right)dt + \hat{g}^{ij}\hat{\beta}_{ij}dW_t^{(ij)}  \right]
	\end{split}
	\end{equation*}
	and so, because  $d\Slog\left(\hat{g}_t^{-1/2}\right) = d\hat{g}_t^{-1/2}\hat{g}_t^{1/2}$, the time evolution is given by
	\begin{equation}\label{fieldtheoryrepr}
	d\Phi_t(\mathbf{x}) = \left( -\frac{i}{\hbar}\hat{H}_{ext} - \frac{1}{2}\hat{g}^{ij}\hat{\alpha}_{ij} + \frac{3}{8} (\hat{g}^{ij}\hat{\beta}_{ij})^2 \right) \Phi_t(\mathbf{x}) dt + \hat{g}^{ij}\hat{\beta}_{ij}\Phi_t(\mathbf{x})dW_t^{(ij)}.  
	\end{equation}
	We are now in the position to discuss the physical relevance of this representation. To keep the discussion simple, we consider the deterministic case only ($\beta_{ij} = 0$). Consider the action of the Schr\"{o}dinger field in flat space. It is given by
	\begin{equation}
	\begin{split}
	S &=  \int dtd^3\mathbf{x}\mathcal{L}(\Phi,\Phi^*,\partial_t \Phi, \partial_t \Phi^*)  \\
	&=\int dtd^3\mathbf{x} \left(i\hbar\Phi^*(\mathbf{x}) \overleftrightarrow{\partial}_t \Phi(\mathbf{x}) - \mathcal{H}[\Phi,\Phi^*] \right),
	\end{split}
	\end{equation}
	where $A\overleftrightarrow{\partial}_tB = (A \partial_tB + \partial_tA B)/2$ is the symmetryzed time-derivative and $\Hi[\Phi,\Phi^*]$ is the Hamiltonian functional. From the Eulero-Lagrange equations for the lagrangian $\mathcal{L}(\Psi,\Psi^*,\partial_t \Psi, \partial_t \Psi^*) $, one obtains the Schr\"{o}dinger equation. The standard way to pass from a flat to a curved manifold is via the \emph{minimal coupling prescription} \cite{gasperini2013theory}. In our case, it consist simply in replacing the flat volume element $d^3\mathbf{x}$ with the curved one $\sqrt{g}d^3\mathbf{x}$ and the partial derivative $\partial_{x_i}$ with the covariant derivative $\nabla_{x_i}$. By doing that, the lagrangian transforms as follows
	\begin{equation*}
	\mathcal{L}_{flat} \mapsto \mathcal{L}_{curved} = \sqrt{g} \left[i\hbar\Phi^*(\mathbf{x}) \overleftrightarrow{\partial}_t \Phi(\mathbf{x}) - \mathcal{H}'[\Phi,\Phi^*]\right],
	\end{equation*}
	where $\mathcal{H'}$ is different from $\mathcal{H}$ since the partial derivatives are replaced by the covariant ones. Employing the lagrangian $\mathcal{L}_{curved}$ in the Eulero-Lagrange equations one obtains exactly the time-evolution \eqref{Es1rep1} in the deterministic case, when the Hamiltonian functional is given by
	\begin{equation*}
	\mathcal{H}'[\Phi,\Phi^*] = \int d^3\mathbf{x}'\int d^3\mathbf{x} \Phi_t(\mathbf{x}')^*H(\mathbf{x}',\mathbf{x})\Phi_t(\mathbf{x}),
	\end{equation*}
	with $H(\mathbf{x}',\mathbf{x})$ equal to the matrix element in the position basis of $\hat{H}_{ext}$, i.e. $\langle \mathbf{x}' |\hat{H}_{ext} |\mathbf{x}\rangle$.
	
	Now consider the representation of the time evolution in $\Hi_K$ given by \eqref{EQrep2}. Since $\hat{h}_t^{\dagger} = \hat{g}_t^{-1/4}$, we have
	\begin{equation}\label{Es1rep2}
	d\chi_t(\mathbf{x}) = -\frac{i}{\hbar} \hat{K} \chi_t(\mathbf{x})dt + d\Slog(\hat{g}_t^{-1/4}) \chi_t(\mathbf{x}).
	\end{equation} 
	Applying the \Ito formula we get
	\begin{equation*}
	\begin{split}
	d \hat{h}_t^{\dagger} &= -\frac{1}{2\sqrt{\hat{g}_t^3}} d\sqrt{\hat{g}_t} + \frac{3}{8\sqrt{\hat{g}_t^5}} d\llbracket \sqrt{\hat{g}_t},\sqrt{\hat{g}_t} \rrbracket  \\
	&= -\frac{1}{4\hat{g}_t} \left[ \left( \hat{g}^{ij} \hat{\alpha}_{ij} - \left(1 + \frac{3}{2\sqrt{\hat{g}_t}}\right)\frac{(\hat{g}^{ij}\hat{\beta}_{ij})^2}{4} \right) dt + \hat{g}^{ij}\hat{\beta}_{ij}dW_t^{(ij)} \right]
	\end{split}
	\end{equation*}
	from which we obtain that
	\begin{equation*}
	d\chi_t(\mathbf{x}) = \left[-\frac{i}{\hbar} \hat{K} -\frac{(\hat{h}_t^{\dagger})^3}{4}\left(\hat{g}^{ij}\hat{\alpha}_{ij}-\left(1+\frac{3(\hat{h}_t^{\dagger})^{2}}{2}\right)\frac{(\hat{g}^{ij}\hat{\beta}_{ij})^2}{4} \right)\right] \chi_t(\mathbf{x})dt+\hat{g}^{ij}\hat{\beta}_{ij}\chi_t(\mathbf{x})dW_t^{(ij)}
	\end{equation*}
	This representation of the time evolution reproduces the results for a quantum particle on a manifold of \cite{dewitt1957dynamical,dewitt1952point}, when the deterministic case is considered. Moreover, the transformed Hamiltonian $\hat{K} = \hat{h}_t^{\dagger}\hat{H}_{ext}\hat{h}_t$ constructed with the $\hat{H}_{ext}$ obtained by field theoretical considerations, corresponds with the Hamiltonian following the prescriptions given in \cite{dewitt1957dynamical}, which makes this representation interesting to describe a quantum particle on a stochastic manifold.
	
	Note that, even in presence of random perturbations of the metric, the time-evolution remains unitary. No Lindblad-type dynamics is obtained for essentially the same reasons discussed in \ref{2.4}, and the probability density must be interpreted as a conditional probability. Let us now relax the condition \eqref{boundedop} to \eqref{relaxedcond} which seems more physical. Indeed, it requires that the volume element $dV_t$ does not explode during the time evolution, but also that the volume element cannot be arbitrarily small. In this sense condition \eqref{relaxedcond} requires the existence of an elementary volume element. As observed at the end of section \ref{derivationtimeevolution}, only the representation \eqref{fieldtheoryrepr} still makes sense, the one which can be also obtained by using field theoretic methods. It is interesting to observe that such elementary volume element is needed in order to describe a quantum particle using the Hilbert space $\Hi_K$, which is the simplest Hilbert space on which such a description may take place.

	\section{Conclusion}
	
	The general case $L_2(\mathcal{M}_t,dm_t)$ can be treated combining the two approaches presented in this article. We want to emphasise again the lack Lindblad-type dynamics in both cases even in presence of stochastic perturbations. This is particularly interesting in light of decoherence and is essentially a consequence of the fact that the observables of the quantum system are stochastic too, since they are themselves defined on a stochastic Hilbert space. Only by neglecting this fact, which means to consider observables that have no stochastic dependence, one can obtain a Lindblad-type dynamics (using for example the cumulant expansion method \cite{kubo1962generalized,breuer2002theory}) typical of an open quantum system. This may give rise to decoherence, which is expected to happen in the position basis. However, neglecting this stochastic dependence does not seem to be justifiable in general, in the kind of problems considered here. 
	
	\section{Acknowledgement}
	
	L.C. thanks L.Asprea, G.Gasbarri and S.Marcantoni for the various discussions had during the elaboration of this work.
	
	\section{Appendix}
	
	\subsection{Unitarity of \eqref{NewTimeEvolution} restricted to $L_2\left( \left[-\frac{1}{2},\frac{1}{2}\right] \right)$ with Dirichlet boundary conditions.}
	
	Let us consider the following quantum dynamics on $L_2(\Rea)$ written in the position representation
	\begin{equation}\label{equaz1}
		i\hbar \frac{d}{dt} \Psi(x,t) = \hat{H}(t,v_0) \Psi(x,t) := \left( -\frac{\hbar^2}{2m}\frac{d^2}{dx^2} + V(x,t) \right)\Psi(x,t),
	\end{equation}
	with some normalized initial condition $\psi(x;0) \in L_2(\Rea)$, where 
	\begin{equation*}
	V(x,t) =
	\begin{cases}
	g(x) &\mbox{ for } x \in [a_t,b_t], \\
	v_0 &\mbox{ for } x \in (-\infty,a_t) \cup (b_t,+\infty),
	\end{cases}
	\end{equation*}
	with $v_0 > 0$ and $g(x)$ is a suitable well behaved function. This describes a finite potential well of height $v_0$, with a non constant term inside the well.
	$a_t$ and $b_t$ are the two extrema of the well and fulfill relations \eqref{BoundarySDE}. As domain of the Hamiltonian operator we use $D(\hat{H}(t,v_0)) = H_2(\Rea)$ for all $t$, on which it is self-adjoint for well behaved functions $g(x)$. Hence, \eqref{equaz1} defines a unitary dynamics on $L_2(\Rea)$.
	We assume that $a_t < 0 < b_t$ without loosing generality. This can be always done performing a suitable (possibly time dependent) translation by some factor $d(t)$, which would affect only the region $[a_t,b_t]$ changing the potential ($g(x) \rightarrow g(x - d(t))$).
	 Let $\psi_{I}(x,t)$, $\psi_{II}(x,t)$ and $\psi_{III}(x,t)$ be the wave function in the regions $(-\infty,a_t)$, $[a_t,b_t]$ and $(b_t,+\infty)$ respectively. To solve \eqref{equaz1} we need to specify the behavior of $\psi(x,t)$ at the boundaries of the well. Requiring continuity of the first derivative we get
	\begin{equation}\label{continuity_conditions}
	\begin{split}
			\psi_I(a_t,t)=\psi_{II}(a_t,t), &\mspace{30mu} \psi_{II}(b_t,t)=\psi_{III}(b_t,t), \\
			\psi_I'(a_t,t)=\psi_{II}'(a_t,t), &\mspace{30mu} \psi_{II}'(b_t,t)=\psi_{III}'(b_t,t).
	\end{split}
	\end{equation}
	These conditions put constraints on the possible allowed energies as in the more traditional case $g(x) = 0$, well studied in standard quantum mechanics textbook.
	In the regions $(-\infty,a_t)$ and $(b_t,+\infty)$, the solution can be written explicitly. In particular, since the overall wave function is square integrable one can easily conclude that for the bounded states ($v_0 > E$)
	\begin{equation}\label{lateralsolution}
		\psi_I(x,t) = c_1(\rho) e^{\rho x-i\frac{E}{\hbar}t}, \mspace{30mu} \psi_{III}(x,t) = c_2(\rho)e^{-\rho x-i\frac{E}{\hbar}t}
	\end{equation}
	where $\rho^2 = 2m(v_0 - E)/\hbar^2$ and $c_1(\rho)$, $c_2(\rho)$ are two constants that may depend of $\rho$. We are interested in studying the limit $v_0 \rightarrow \infty$, i.e. $\rho \rightarrow \infty$. To get some insight on the possible asymptotic behavior when $\rho \rightarrow \infty$, we consider \eqref{continuity_conditions}, and in particular we write
	\begin{equation}
		\frac{1}{\rho} = \frac{\psi_I(a_t,t)}{\psi_I'(a_t,t)} = \frac{\psi_{II}(a_t,t)}{\psi_{II}'(a_t,t)}.
	\end{equation}
	For $\rho \rightarrow \infty$, the LHS goes to zero while in the RHS we may have two possibilities: $\psi_{II}(a_t,t) =  0$ or $\psi_{II}'(a_t,t) = \infty$ in the limit. However this last case is excluded since $\psi_{II} \in H_2(\Rea)$ for any value of $\rho$. Similar considerations hold for the second set of conditions in $b_t$. Hence we conclude that when $\rho \rightarrow \infty$
	\begin{equation}
		\psi_{II}(a_t,t) =  0 \mspace{30mu} \psi_{II}(b_t,t)=0.
	\end{equation}
	This also mean that $\psi_I(a_t,t) = 0$ and $\psi_{III}(b_t,t) = 0$, and so the constants $c_1(\rho)$ and  $c_2(\rho)$ remain finite in the limit (or at most diverge slower than an exponential function). This means that in the limit $\rho \rightarrow \infty$ $\psi_I(x,t) = 0$ and $\psi_{III}(x,t) = 0$ (hence in the limit $\hat{H}(t,v_0)$ acts as \eqref{ExtendedHamiltonian}).
	This shows that for any initial condition in $L_2(\Rea)$, in the limit $v_0\rightarrow\infty$, the time-evolution induced by \eqref{equaz1} gives functions belonging to $L_2([a_t,b_t])$ vanishing at the boundaries, i.e. it reduces to the time-evolution induced by equation \eqref{SchEq}.
	
	At this point we use the operator $\hat{W}$, defined in \eqref{Transformation}, to map the previous dynamics, where the well in the potential is time dependent, into an analogous problem where the well of the potential is fixed for any value of $v_0$. The unitary time-evolution in this case is given by
	\begin{equation}
		\hat{V}_t(v_0) := \hat{W}e^{-\frac{i}{\hbar}\hat{H}(v_0),t}\hat{W}^{\dagger}.
	\end{equation}
	From the previous discussion, we know that as $v_0 \rightarrow \infty$ the operator $\hat{V}_t(v_0) \rightarrow \hat{V}_t$, where $\hat{V}_t$ is the linear time-evolution induced by \eqref{NewTimeEvolution} with Dirichlet boundary conditions at the fixed boundaries $\pm 1/2$. More precisely
	\begin{equation*}
		\lim_{v_0 \rightarrow \infty} \hat{V}_t(v_0)|\Phi_0\rangle = \hat{V}_t|\Phi_0\rangle
	\end{equation*}
	where $|\Phi_0\rangle = \hat{W}|\Psi_0\rangle$, i.e. the convergence is in the strong sense. $|\Psi_0\rangle \in L_2(\Rea)$ is the initial condition and we can choose all the initial conditions with support on $\left[-\frac{1}{2},\frac{1}{2}\right]$, i.e. elements of $L_2\left( \left[-\frac{1}{2},\frac{1}{2}\right] \right)$. At this point we have that
	\begin{equation*}
		\hat{V}_t^\dagger \hat{V}_t = \lim_{v_0 \rightarrow \infty} \hat{V}_t(v_0)^\dagger\hat{V}_t(v_0) = \Id.
	\end{equation*}
	In the same way one proves that $\hat{V}_t\hat{V}_t^\dagger = \Id$, showing that $\hat{V}_t$ is unitary operator also when restricted to the Hilbert space $L_2\left( \left[-\frac{1}{2},\frac{1}{2}\right] \right)$.

	\subsection{The scalar product on $\Hi_K$.}
	
	Let us prove that \eqref{elemtaryscalarp} is a scalar product on $\Hi_K$. Taken $\phi,\psi \in \Hi_K$, by definition of $\Hi_K$ one has that $\psi = \psi_t \in L_2(\mathcal{M},m_t(x)dx)$ and $\phi = \phi_s \in L_2(\mathcal{M},m_s(x)dx)$ for some $s,t \in \Rea$. By construction \eqref{elemtaryscalarp} is linear in the second entry. Moreover $\langle \psi | \phi \rangle =  [\langle \phi | \psi \rangle]^*$. Indeed,
	\begin{equation*}
	\begin{split}
	[\langle \phi | \psi \rangle]^* &= [\langle \phi_t| \hat{h}_t^\dagger \hat{h}_s \psi_s\rangle_t]^* 
	= \langle \psi_s\hat{h}_s \hat{h}_t^\dagger |\phi_t\rangle_t \\
	&= \langle \psi_s |\hat{h}_s^\dagger \hat{h}_t |\phi_t\rangle_s = \langle \psi|\phi \rangle.
	\end{split}
	\end{equation*}
	Since \eqref{elemtaryscalarp} reduces to the ordinary inner product on $L_2(\mathcal{M},m_t(x)dx)$, positivity and non-degeneracy follows.
	
	\subsection{Forms of $\hat{h}_t$, $\hat{h}_t^\dagger$ and $\hat{m}_t^{\dagger}$}
	
	Let us find the expression of $\hat{h}_t^{\dagger}$ and $\hat{h}_s$ introduced in section \ref{hhdagasection}. From \eqref{measureoperator} and because $\hat{h}_t = \sqrt{\hat{m}_t}$, we can see that
	\begin{equation*}
	\hat{h}_t \psi_t(x) = \sqrt{m_t(x)}\psi_t(x).
	\end{equation*}
	Clearly if $\psi_t(x) \in L_2(\mathcal{M},m_t(x)dx)$, then $\sqrt{m_t(x)}\psi_t(x) \in L_2(\mathcal{M},dx)$. To define $\hat{h}_t^\dagger$ one can use $\hat{h}_t^\dagger\hat{h}_t = \Id$ (the identity is in $L_2(\mathcal{M},m_t(x)dx)$), obtaining
	\begin{equation*}
	\begin{split}
	\langle \psi_t | \hat{h}_t^\dagger\hat{h}_t \phi_t \rangle_t &= \int_\mathcal{M} \psi_t(x)^* \hat{h}_t^\dagger\hat{h}_t\phi_t(x)m_t(x)dx \\
	&= \int_\mathcal{M} \psi_t(x)^* \hat{h}_t^\dagger\sqrt{m_t(x)}\phi_t(x)m_t(x)dx = \langle \psi_t | \phi_t \rangle_t,
	\end{split}
	\end{equation*}
	from which we deduce that
	\begin{equation}\label{hdaga}
	\hat{h}^{\dagger}_t \tilde{\psi}_t(x) = \frac{1}{\sqrt{m_t(x)}}\tilde{\psi}_t(x).
	\end{equation}
	From this, it is possible to see that $[\hat{m}_t]^{\dagger} = \hat{m}_t^{-1}$ when $\hat{m}_t$ is considered as an operator acting on $\Hi_K$.
	
	\subsection{The stochastic logarithm}
	
	Let $Y_t$ be a stochastic process on a probability space $\PS$. Assume that $Y_t$ admits stochastic differential and remains positive for any $t$. The \emph{stochastic logarithm} of $Y_t$, i.e. $\Slog(Y_t)$, is defined as the unique process fulfilling the SDE \cite{klebaner2005introduction}
	\begin{equation*}
	\begin{cases}
	d \Slog(Y_t) = \frac{dY_t}{Y_t}, \\
	\Slog(Y_0) = 0.
	\end{cases}
	\end{equation*}
	If $X_t$ is another stochastic process on $\PS$ admitting stochastic differential, the stochastic logarithm of the product $X_tY_t$ is equal to
	\begin{equation}\label{stochasticlogprod}
	d \Slog(X_t Y_t) = d \Slog(X_t) + d \Slog(Y_t) + \frac{d\llbracket X_t,Y_t \rrbracket}{X_tY_t}
	\end{equation}
	To see that, it is enough to apply the chain rule of stochastic calculus. Indeed
	\begin{equation*}
	\begin{split}
	d \Slog(X_t Y_t) &= \frac{d(X_t Y_t)}{X_tY_t} =	\frac{dX_t Y_t + X_tdY_t + d \llbracket X_t,Y_t \rrbracket}{X_t Y_t} \\
	&= \frac{dX_t}{X_t} + \frac{dY_t}{Y_t} + \frac{d\llbracket X_t,Y_t\rrbracket}{X_tY_t},
	\end{split}
	\end{equation*}
	from which one obtains \eqref{stochasticlogprod}. In particular, in the case $X_t = Y_t^{-1}$, \eqref{stochasticlogprod} implies that
	\begin{equation}\label{inverseslog}
	d\Slog(Y_t^{-1}) = - d\Slog\left( Y_t \right) - d\llbracket Y_t^{-1},Y_t \rrbracket.
	\end{equation}
	Note the difference with the ordinary logarithm, where $d\log(x^{-1}) = -d\log(x)$.

	\bibliographystyle{unsrt}
	\bibliography{bibliouv}
	
\end{document}